\documentclass{ws-ijmpa}

\begin{document}

\markboth{X. J. Wen et al.} {Color-flavor locked strangelets in a
quark mass density-dependent model}

%
\catchline{}{}{}{}{}
%

\title{Color-flavor locked strangelets in a quark mass density-dependent model}

\author{X.\ J.\ Wen,$^{1,2,3}$\footnote{ wenxj@ihep.ac.cn} \ G.\ X.\ Peng, $^{2,1}$ P.\ N.\ Shen$^{1,2}$}

\address{\footnotesize \sl $^1$Institute of High Energy Physics, Chinese
Academy of Sciences, Beijing 100049, China \\
\footnotesize \sl $^2$China Center of Advanced Science and
Technology
   (World Lab.), Beijing 100080, China\\
\footnotesize \sl $^3$Graduate University of Chinese Academy of
Sciences, Beijing 100049, China}


\maketitle
\begin{history}
\end{history}

\begin{abstract}
The color-flavor locked (CFL) phase of strangelets is investigated
in a quark mass density-dependent model. Parameters are determined
by stability arguments. It is concluded that three solutions to the
system equations can be found, corresponding, respectively, to
positively charged, negatively charged, and nearly neutral CFL
strangelets. The charge to baryon number of the positively charged
strangelets is smaller than the previous result, while the charge of
the negatively charged strangelets is nearly proportional in
magnitude to the cubic-root of the baryon number. However, the
positively charged strangelets are more stable compared to the other
two solutions.

\keywords{Quark matter; Color-flavor locking; Strangelets.}
\end{abstract}

\ccode{PACS numbers: 24.85.+p, 12.38.Mh, 25.75.-q}

\section{Introduction}
\label{sec:intro}

At high densities, quark matter, or its finite lumps, the so-called
strangelets, may form the color superconductor by chiral-symmetry
violating condensate.\cite{Rajagopal2000,Alford2001} Most probably
strange quark matter (SQM) is more stable than the hadronic
matters,\cite{JaffePRD1984} which may have far-reaching consequences
astronomically and
cosmologically.\cite{BanksPRD2005,Khlopov,khlopov1} It has potential
applications to the astrophysics of compact
stars,\cite{Alford1991,Alfordjhep,Alford2004,LugonesAA2003,Lugones2004,Lugones2005}
and to high energy heavy ion collisions.\cite{GyulassyNPA2005}
Recently, the properties of CFL phase have been investigated by many
authors with various models
.\cite{MAlfordNPB99,MadsenPRL2001,Huangm03PRD67,Shovgovy2003,Mishra2005,Kiriyama2005,Hou2004,XBZhangPRD2004,huazhong}
Rajagopal and Wilczek demonstrated that the CFL quark matter is
automatically neutralized without any requirements
 on electrons.\cite{KRajagopalPRL2001} Madsen studied the positively
charged CFL strangelets, and found that the CFL strangelets may be
more stable than the ordinary strangelets without color-flavor
locking.\cite{MadsenPRL2001,MadsenJPG2002} Alford investigated the
new gapless CFL phase in neutral cold quark
matter.\cite{MAlfordPRL04,MAlfordPRD2005} Lugones and Horvath
derived an approximate equation of state (EOS) to lowest order in
$\Delta$ and \textit{m} and found the effects of pairing on EOS and
the window of stability for CFL strange matter.\cite{LugonesPRD2002}

In the standard MIT bag model, quark masses are constant. As is well
known, however, the quark mass changes with environment, or, in
other words, it depends on density. Such masses are usually called
effective masses.
\cite{henleyNPA1990,BrownPRL1991,CohenPRL1991,Cohen1992,waleckaNP1995}
Effective masses and effective bag constants for quark matter have
been extensively discussed within the Nambu--Jona-Lasinio
model.\cite{BuballaPLB1999} In recent years, the quark mass
density-dependent model (QMDD) has been successful in describing the
unpaired quark
matter.\cite{Fowler1981,Benrenuto1995,Lugones1995,PengPRC59,Peng00PRC61,Peng00PRC62,Lugones2003qmdd,Zhang2001,Zhang2002,Zhangprc2003,Zhang2003,PRC72(2005)015204}
Very recently, we also studied CFL strangelets within the framework
of bag model.\cite{PengPLB2006} Now in this paper, we study how the
density dependence of quark masses influences the properties of CFL
strangelets. It is found that for a given baryon number, there are
also three kinds of CFL strangelets which are, respectively,
charge-positive (slet-1), negative (slet-2), and nearly neutral
(slet-3). These multiple solutions are determined by the multiple
values of quarks Fermi momenta satisfying the self-consistent
condition. The finite-size effects of strangelets with fixed Fermi
momenta influence the fractions of quarks as well as the charge of
strangelets. For slet-1, the ratio of the charge to squared
cubic-root baryon number is only half of that in the pure bag model,
while the charge of slet-2 is proportional, in magnitude, to the
cubic-root of baryon number. For the same parameters, the slet-1 is
more stable.

This paper is organized as follow. In the subsequent
Sec.~\ref{Sec:thermo}, we briefly give the thermodynamic formulas
used for the study of CFL strangelets in the mass density-dependent
model. Then we present the numerical results and discussions in
Sec.~\ref{Sec:numerical}. The Sec.~\ref{Sec:conls} is a short
summary.
\section{Thermodynamic formulas and mass density dependence of quark masses}
\label{Sec:thermo}

In this paper, the CFL phase consists of up, down, and strange
quarks in three-color QCD.  The symmetrical color-flavor locked
phase needs an attractive interaction between two quarks near the
Fermi surface with equal and opposite momentum. The Fermi surface is
vividly fixed in the momentum space. The condensate of Cooper pairs,
associated with the group of $ud$, $us$ and $ds$, spontaneously
breaks the color gauge symmetry. At high densities, quarks of three
colors and three flavors are allowed to pair and have the same Fermi
momentum.

In the pairing ansatz\cite{MAlfordNPB99}
\begin{equation}
\langle \psi^\alpha_a C\gamma_5 \psi^\beta_b \rangle \sim \Delta_1
\epsilon^{\alpha\beta 1}\epsilon_{ab1} \!+\! \Delta_2
\epsilon^{\alpha\beta 2}\epsilon_{ab2} \!+\! \Delta_3
\epsilon^{\alpha\beta 3}\epsilon_{ab3} \,, \label{condensate}
\end{equation}
we consider the case $\Delta_3\approx\Delta_2=\Delta_1=\Delta$ for
common CFL phase. Here $\psi^\alpha_a$ is a quark field with color
$\alpha=(r,g,b)$ and flavor $a=(u,d,s)$. The thermodynamic potential
of the color-flavor locked strangelets can be written as
\cite{MadsenPRL2001,KRajagopalPRL2001,MAlfordPRD2001}
\begin{eqnarray}\Omega=\sum_i\Omega_{\mathrm{f},i}+\Omega_{\mathrm{pair},V}+B\,,\label{Eq:omega}
\end{eqnarray}
where the bag constant $B$ is appended as normally done.
Eq.~(\ref{Eq:omega}) is derived from microscopic models under the
condition that $\Delta/\mu$ is small.\cite{MAlfordPRD2001} For
pairing contribution, we include the volume term
$\Omega_{\mathrm{pair},V}\approx -3\Delta^2\bar{\mu}^2/\pi^2$, which
will appear in the number density, energy and pressure \textit{et}c.
The finite-size contribution to $\Omega_\mathrm{pair}$ is assumed
small and reasonably neglected.\cite{MadsenPRL2001} The quantity
$\bar{\mu}\equiv(\mu_u+\mu_d+\mu_s)/3$. $\Omega_{\mathrm{f},i}$
denotes the normal quark contribution from flavor type $i$ ( $i=u,
d, s$ ), i.e.,
\begin{equation}
 \Omega_{\mathrm{f},i}=-\ T\int_0^{p_\mathrm{F}} \ln\left\{1\pm
\exp\left[-\frac{\varepsilon_i(p_i)-\mu_i}{T}\right]\right\}
n^\prime_i(p, m_i, R) d p\,.
\end{equation}
The integral upper limit $p_\mathrm{F}$ means Fermi momentum, which
is equal for $u, d, s$ quarks in CFL quark
matter.\cite{KRajagopalPRL2001} When $T=0$, this expression can be
simplified, giving
\begin{eqnarray} \Omega_{\mathrm{f},i}=\int_0^{p_\mathrm{F}}
(\varepsilon_i-\mu_i)n^\prime_i(p, m_i, R) dp\,,\label{omei}
\end{eqnarray}
where the density of states for strangelets from the
multi-reflection theory\cite{R.Balian} is
\begin{equation}
n^\prime_i(p, m_i,
R)=g_i\left[\frac{1}{2\pi^2}p^2+\frac{S}{V}f^s_i\left(\frac{p}
{m_i}\right)p+\frac{C}{V}f^c_i\left(\frac{p}{m_i}\right)\right].
\end{equation}
Here $V=\frac{4}{3}\pi R^3$, $S=4\pi R^2$, and $C=8\pi R$. The
function $f_i^S$ and $f_i^C$ are respectively referred as
 surface term \cite{JaffePRD1984,Berger1987} and curvature term \cite{Madsen1993,Madsen1994}
 and given by
\begin{eqnarray}
f^S_i\left(\frac{p}{m_i}\right)&=&-\frac{1}{4\pi^2}\cot^{-1}\left(\frac{p}{m_i}\right),\\
f^C_i\left(\frac{p}{m_i}\right)&=&\frac{1}{12\pi^2}\left[1-\frac{3p}{2m_i}
 \cot^{-1}\left(\frac{p}{m_i}\right)\right].
\end{eqnarray}

For convenience, we define a parameter
$\phi_i\equiv\arctan(p_\mathrm{F}/m_i)$. If the thermodynamic
potential density in Eq.~(\ref{omei}) is divided into three parts
i.e., $\Omega_{\mathrm{f},i}=\Omega^V_i+\frac{3}{R}\Omega^S_i +
\frac{6}{R^2}\Omega^C_i$, then the volume term $\Omega^V_i$, the
surface term $\Omega^S_i$, and the curve term $\Omega^C_i$ can have
analytical forms as,

\begin{eqnarray}
\Omega^V_i&=&-\frac{g_im_i^4}{16\pi^2}\left\{\frac{8}{3}\frac{\mu_i}{m_i}\tan^3\phi_i+\ln\left[\tan\phi_i+\sec\phi_i\right]
\right.\nonumber\\
&&\left.\phantom{-\frac{g_im_i^4}{16\pi^2}\{} - \tan\phi_i\sec\phi_i^3 - \tan^3\phi_i \sec\phi_i \right\}\,, \label{Eq:omegaV}\\
\Omega^S_i&=&-\frac{g_im_i^3}{24\pi^2}\left\{ \ln\left[
\tan\phi_i+\sec\phi_i\right]-\frac{3\mu_i}{m_i}\left[
(\frac{\pi}{2}-\phi_i)\tan^2\phi_i+ \tan\phi_i-\phi_i\right]
\right.\nonumber \\
&&\left.\phantom{-\frac{g_im_i^3}{24\pi^2}\{} +\sec\phi_i
\left[\tan\phi_i
+\sec^2\phi_i(\pi-2\phi_i) \right] -\pi\right \}\,, \label{Eq:omegaS}\\
\Omega^C_i&=&\frac{g_im_i^2}{48\pi^2} \left\{
\ln\left[\tan\phi_i+\sec\phi_i\right]-
\frac{3\mu_i}{m_i}\left[\frac{1}{3}\tan\phi_i
-\tan^2\phi_i(\frac{\pi}{2}-\phi_i)
+\phi_i\right]\right.\nonumber\\
&&\left.\phantom{\frac{g_im_i^2}{48\pi^2}\{}+\tan\phi_i \sec\phi_i-2
\sec^3\phi_i(\frac{\pi}{2}-\phi_i)+\pi \right\}\,.\label{Eq:omegaC}
\end{eqnarray}

The common fermi momentum $p_\mathrm{F}$ is a fictional intermediate
parameter, which can be determined by minimizing the thermodynamic
potential $\Omega$, i.e.,
\begin{eqnarray} \label{con3} \frac{\partial \Omega}{\partial
p_\mathrm{F}}=0\,,
\end{eqnarray}
or, explicitly,
\begin{eqnarray}
\sum_{i=u,d,s}n^\prime_i(p_\mathrm{F},m_i,R) \left[
\sqrt{p_\mathrm{F}^2+m_i^2}-\mu_i \right]=0\,.\label{Eq:chemical}
\end{eqnarray}

By differentiating $\Omega_{\mathrm{f},i}$ with respect to the
chemical potential $\mu_i$ we obtain
\begin{eqnarray}
n_{\mathrm{f},i}&=&g_i p_\mathrm{F}^3/(6\pi^2)+\frac{3g_i
m_i^2}{8\pi^2R} \left[\phi_i-\tan\phi_i -
(\frac{\pi}{2}-\phi_i)\tan^2\phi_i
\right] \nonumber\\
&& + \frac{3g_im_i}{8\pi^2R^2} \left[\phi_i +
\frac{1}{3}\tan\phi_i-(\frac{\pi}{2}-\phi_i)\tan^2\phi_i \right]\,.
\end{eqnarray}
Hence the number density for flavor $i$ is
\begin{eqnarray}n_i=n_{\mathrm{f},i}+\frac{2}{\pi^2}\Delta^2\bar{\mu}\,.\label{Eq:density}
\end{eqnarray}
The second term is from the paring effect. We should accordingly
note that the density depends directly on the chemical potential
$\mu$ and the paring parameter $\Delta$, not merely the Fermi
momentum ``$p_{\mathrm{F}}$''. The corresponding energy density can
then be expressed as
\begin{eqnarray}
E &=&\Omega + \sum_i\mu_in_i \nonumber\\
&=& \sum_i \left[ \Omega_{\mathrm{f},i}+ \mu_i
(n_{\mathrm{f},i}+n_{\mathrm{pair},V})
\right]+\Omega_{\mathrm{pair},V} +B\,.
\end{eqnarray}

In the conventional bag model, quark masses are constant. As is well
known, however, quark masses vary with environment. In fact, not
only quark masses will change but also the coupling constant will
run in the medium.\cite{pertQCD} Effective masses and effective bag
constants for quark matter have been extensively discussed, e.g.,
within the Nambu--Jona-Lasinio model\cite{BuballaPLB1999} and within
a quasiparticle model.\cite{ScherNPA1997} Therefore, in recent
years, the quark mass density-dependent model has been shown to be
successful in the study of quark matter.
\cite{Fowler1981,Benrenuto1995,Lugones1995,PengPRC59,Peng00PRC61,Peng00PRC62,Lugones2003qmdd,Zhang2001,Zhang2002,Zhangprc2003,Zhang2003,PRC72(2005)015204}
The question now is how to parameterize the density dependence of
quark masses. In principle, it should be treated dynamically and be
consistent with the overall chiral symmetries of QCD. In studying
the unpaired phase of quark matter,
\cite{Peng00PRC61,Peng00PRC62,Lugones2003qmdd,PRC72(2005)015204} the
equivalent quark mass is
\begin{eqnarray}
 m_q&=&m_{q0}+\frac{D}{n_\mathrm{b}^{1/3}} \,.
\end{eqnarray}

The baryon number $A$ is connected to the quark number $N_u, N_d$
and $N_s$ by
\begin{eqnarray}\label{con1} A=\frac{1}{3}(N_u+N_d+N_s) \,,
\end{eqnarray}
and the chemical equilibrium requires\cite{JaffePRD1984}
\begin{eqnarray}
\mu_d=\mu_s=\mu_u+\mu_e \label{con2} \,.
\end{eqnarray}
When the baryon number $\textit{A} \ll 10^7$, electrons can't exist
in a strangelet because the electron Compton wavelength exceeds the
sphere radius \textit{R}. So the electron ``effective" chemical
potential is zero in strangelets,  which is consistent with the
viewpoints of Ref.~\refcite{lecture2002} and \refcite{WeberPPNP2005}
to some extent. Figure 11.1 of Ref.~\refcite{lecture2002} and Figure
3 of Ref.~\refcite{WeberPPNP2005} point that when the mass of quark
matter is less than $10^9$ GeV, the electron could be outside the
quark bag. So when $\mu_e=0$, Eq.~(\ref{con2}) becomes
$\mu_u=\mu_d=\mu_s$. With this relation, the chemical potential can
be obtained from Eq.~(\ref{Eq:chemical}) for a fixed $A$ and a given
radius $R$.

The radius of a strangelet with a given baryon number is determined
by minimizing the thermodynamic potential with respect to the
radius,
\begin{eqnarray} \label{con4}\frac{\partial \Omega}{\partial R}=0 \,.
\end{eqnarray}
Or, equivalently, by setting the pressure
\begin{eqnarray}
P=-\Omega- \frac{R}{3}\frac{\partial\Omega}{\partial
R}+n_\mathrm{b}\frac{\partial m_i}{\partial
n_\mathrm{b}}\frac{\partial \Omega}{\partial m_i}-B \label{press}
\end{eqnarray}
to zero, i.e., $P=0$. The first two terms on the right hand side of
Eq.~(\ref{press}) is normal, while the third term is special when
quark masses are density-dependent. As has been shown in literature,
this extra term is necessary to satisfy the basic thermodynamic
requirement, i.e., the energy minimum must be exactly located at the
zero pressure.\cite{Peng00PRC62,PRC72(2005)015204}

 According to Eqs.~(\ref{Eq:omegaV}), (\ref{Eq:omegaS}), and (\ref{Eq:omegaC}), the
derivative of thermodynamic potential density with respect to mass
can be divided into three parts as,

\begin{eqnarray}\frac{\partial\Omega_V}{\partial m_i}&=&\frac{g_i
m_i^3}{4 \pi^2} \Big\{ \tan\phi_i \sec\phi_i- \ln(\tan\phi_i+
\sec\phi_i )\Big\},
\end{eqnarray}
\begin{eqnarray}\frac{\partial \Omega_S}{\partial
m_i}&=&-\frac{g_im_i^2}{8\pi^2}\bigg\{\tan\phi_i \sec\phi_i-2
\phi_i \sec\phi_i + \ln(\tan\phi_i+\sec\phi_i) \nonumber\\
&& \phantom{-\frac{g_im_i^2}{8\pi^2}\{}-(1-\sec\phi_i)\pi
+2\frac{\mu_i}{m_i}[\phi_i-\tan\phi_i]\bigg\}\,,
\end{eqnarray}
\begin{eqnarray}
\frac{\partial \Omega_C}{\partial m_i}&=&\frac{g_i m_i}{96\pi^2}
\bigg\{ 4\ln(\tan\phi_i+ \sec\phi_i)
+2\tan\phi_i\Big(\frac{3\mu_i}{m_i}-2 \sec\phi_i\Big)   \nonumber\\
&&\phantom{\frac{g_i m_i}{96\pi^2}\{} +\pi \Big(2\tan^2\phi_i
\sec\phi_i -4 \sec\phi_i
-\frac{3\mu_i\tan^2\phi_i}{m_i}+4 \Big) \nonumber \\
&& \phantom{\frac{g_i m_i}{96\pi^2}\{} -2\phi_i \Big( 2\tan^2\phi_i
\sec\phi_i +\frac{3\mu_i}{m_i} -4 \sec\phi_i
-\frac{3\mu_i\tan^2\phi_i}{m_i} \Big) \bigg\}.
\end{eqnarray}

Because no electrons are included in the system, finite-size CFL
strangelets are charged. The net electric charge is
$Z=\frac{2}{3}n_u-\frac{1}{3}n_d-\frac{1}{3}n_s$ in unit of the
electric charge of an electron. In principle, the Coulomb effect
should be included, though it is small compared with the strong
interaction. We include it in the numerical calculations.

For the charged CFL strangelets, the Coulomb energy is,
\begin{equation}E_\mathrm{Coul}=\frac{1}{10}\frac{\alpha Z_V^2}{R}+\frac{1}{2}\frac{\alpha
Z^2}{R}\,,
\end{equation}
where the fine structure constant is $\alpha\approx1/137$ and $Z_V$
is the volume term of the total electric charge $Z$ of the CFL
strangelets.

\section{Numerical results and discussions}
\label{Sec:numerical}
We now discuss the parameters adopted in our calculations. The
current quark masses are taken to be $m_{u0}=5$ MeV, $m_{d0}=10$ MeV
$m_{s0}=120$ MeV except for specific indications somewhere. The
super-conducting gap varies from several tens to several hundreds of
MeV in previous papers.\cite{Rapp1998PRL81,Alford1998,Berges1999} In
this paper we take the value $\Delta=100$ MeV, as in
Ref.~\refcite{MadsenPRL2001}.
\begin{figure}[htbp]
\begin{minipage}[t]{0.48\linewidth}
\centering
\includegraphics[width=6cm,height=6cm]{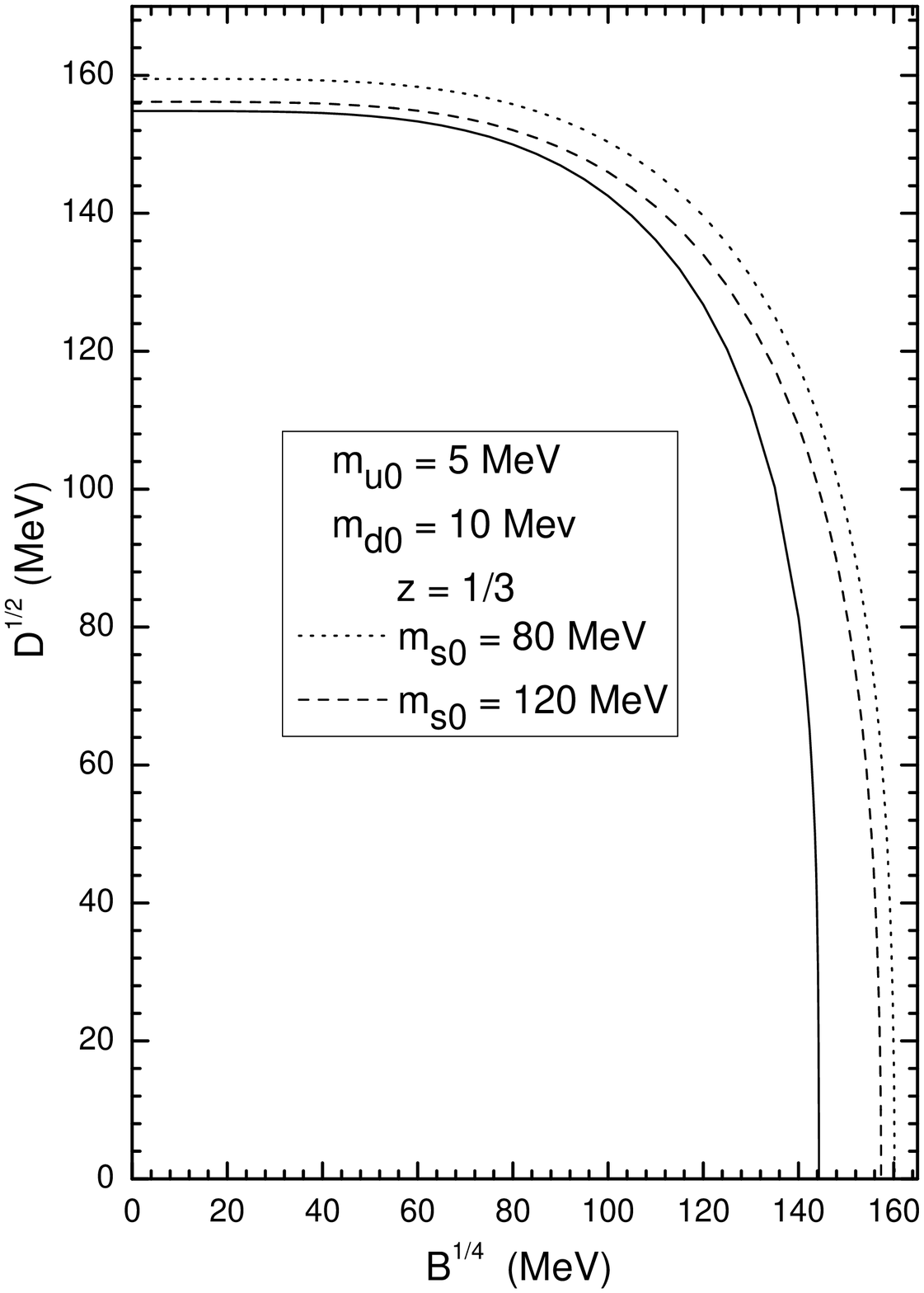}
\caption{The stability window of $D^{1/2}$ and $B^{1/4}$. We take
$m_{u0}=5$ MeV and $m_{d0}=10$ MeV. The lower bound is given by the
full line, while the upper bound is given by the dashed line for
$m_{s0}$ = 80 MeV and by the dotted line for $m_{s0}$ = 120 MeV.
}\label{fig:Dwindow}
\end{minipage}
\hfill
\begin{minipage}[t]{0.48\linewidth}
\centering
\includegraphics[width=6cm,height=6cm]{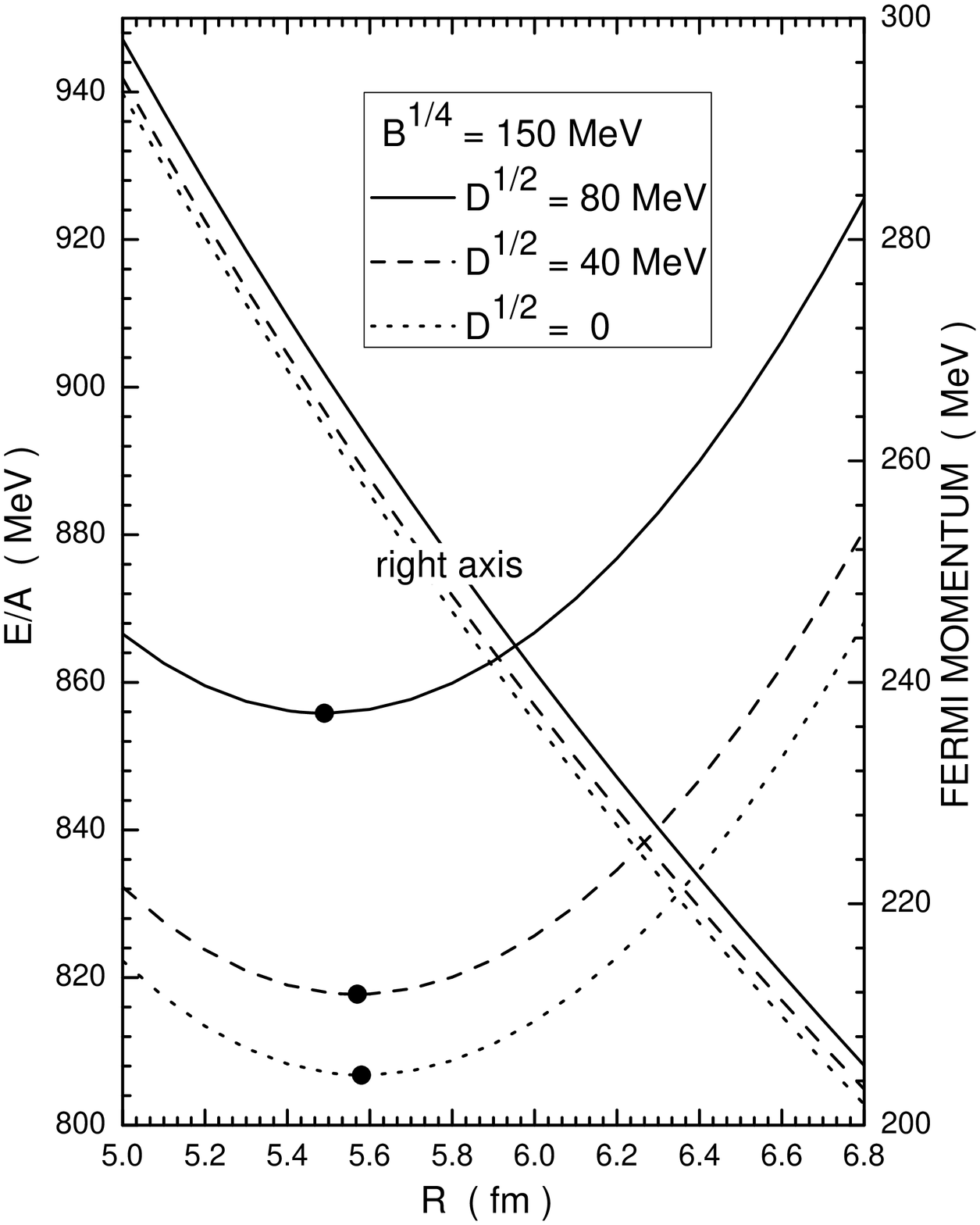}
\caption{Energy per baryon of the positively charged CFL strangelets
for $A=200$. The solid, dashed, and dotted curves stand for
$D^{1/2}=80$ MeV, 40 MeV and 0, respectively. The bag constant is
$B^{1/4}$ = 150 MeV. Fermi momenta are also indicated on the right
axis.}\label{fig:deffect}
\end{minipage}
\end{figure}

\begin{figure}[htbp]
\begin{minipage}[t]{0.48\linewidth}
\centering
\includegraphics[width=6cm,height=6cm]{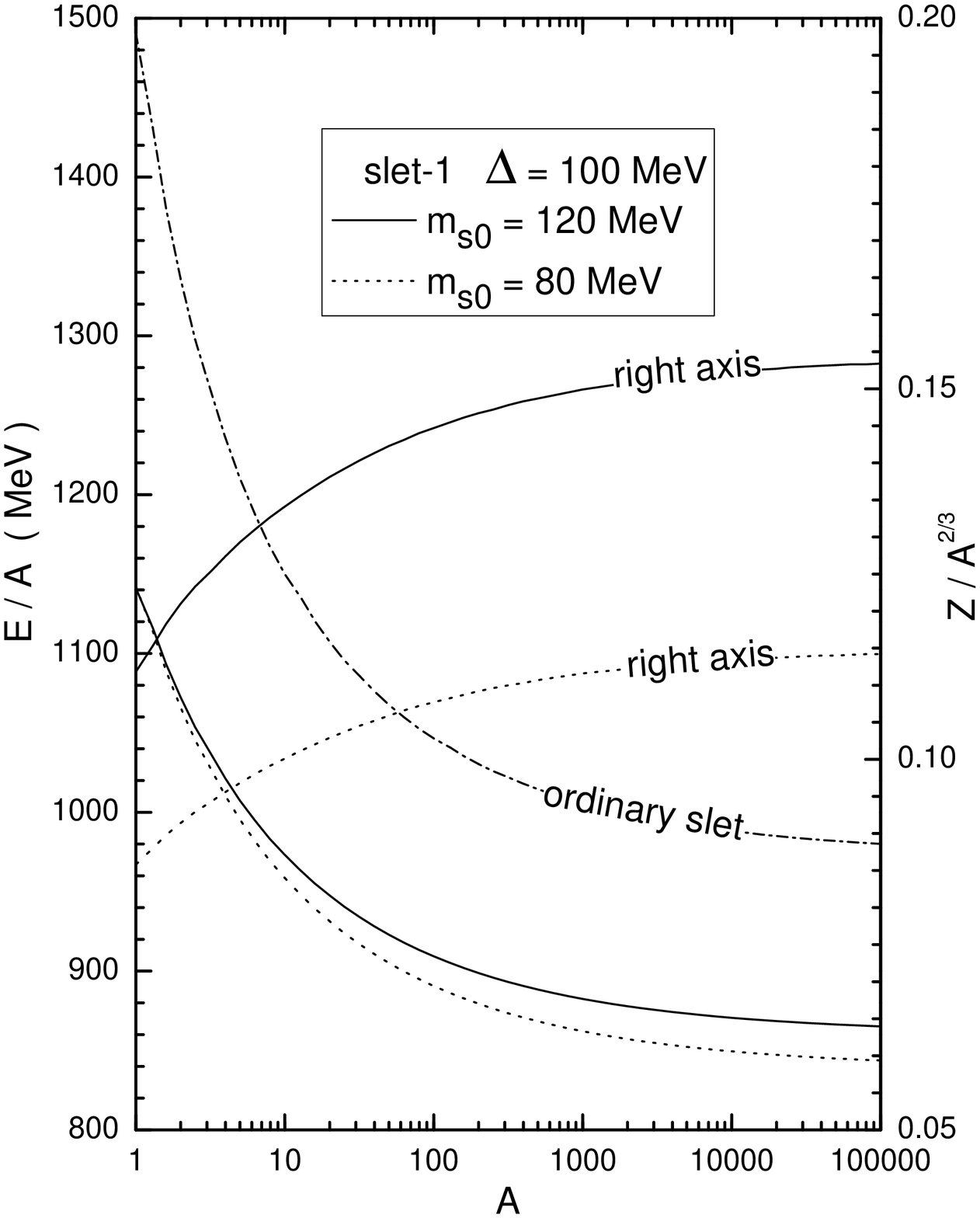}
\caption{Energy per baryon  and electric charge per $A^{2/3}$ as
functions of baryon number $A$ for CFL strangelets (slet-1) are
shown. Here Fermi momentum $p_\mathrm{F}$ is in the usual range of
$250\sim300$ MeV. The ordinary strangelets are also
plotted.\label{fig:033eaza}}
\end{minipage}
\hfill
\begin{minipage}[t]{0.48\linewidth}
\centering
\includegraphics[width=6cm,height=6cm]{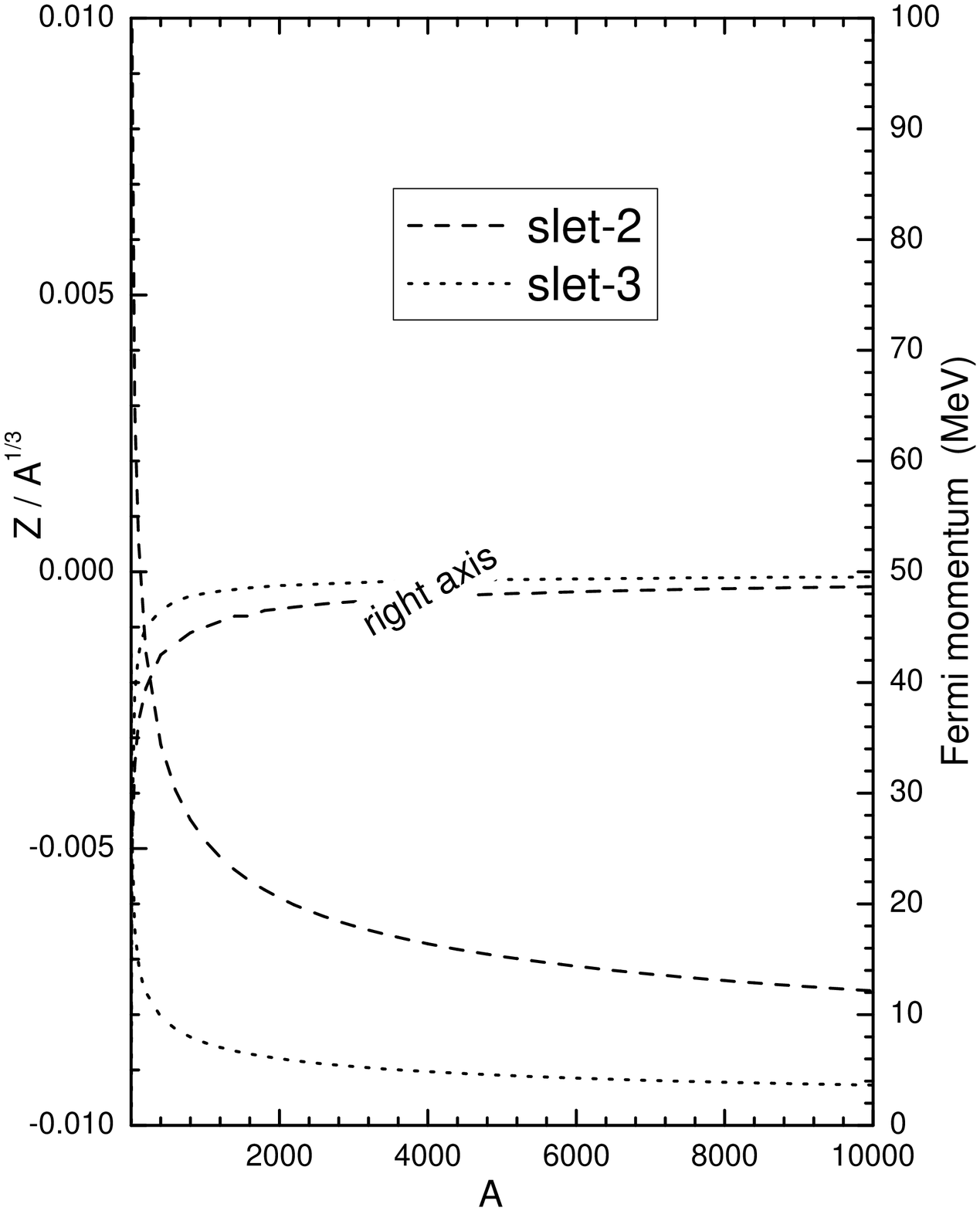}
\caption{ Charge of strangelets for slet-2 (dashed lines) and slet-3
(dotted lines). The slet-3 is nearly neutral, while negatively
charged slet-2 is proportional to $A^{1/3}$. the corresponding Fermi
momenta are also indicated on the right axis.
}\label{fig:slet2slet3}
\end{minipage}
\end{figure}

It is known that the lowest energy per baryon in ordinary nuclei is
930 MeV for iron. In order not to contradict with standard nuclear
physics, the energy per baryon at zero temperature should be greater
than 930 MeV for two flavor quark matter, and less than 930 MeV for
three flavor quark matter so that SQM can have a chance to be
absolutely stable. We show the stability window in
Fig.~\ref{fig:Dwindow}. The horizontal axis, namely $D=0$ and
constant quark mass $m_q=m_{q0}$ ($q=u,d,s$), shows a range of
(144.297,~157.3634) MeV for the constant $B^{1/4}$ in bag model. If
we require the energy per baryon less than the mass of nucleons
$E/n_\mathrm{b}=939$ MeV, we can derive the same (meta)stable range
of $B$ as the result in Ref.~\refcite{JaffePRD1984}. The vertical
axis shows the previous range of (154.8278,~156.1655) MeV for
$\sqrt{D}$ in QMDD model.\cite{PRC72(2005)015204} The parameter
pairs ($D^{1/2}$, $B^{1/4}$) under the solid line or above the
dashed line will not be adopted for stability purpose. If
$B^{1/4}=150$ MeV, a range for $D$ can be found.  For baryon number
$A$=200, the three lines of energy per baryon are shown respectively
for $D^{1/2}$=80, 40 and 0 in Fig.~\ref{fig:deffect}. The full dots
denote the zero pressure points, which are exactly the minima of the
energy per baryon. The corresponding Fermi momenta, indicated on the
right axis, are also affected by the mass parameter $D$.

\begin{figure}[h]
\epsfig{file=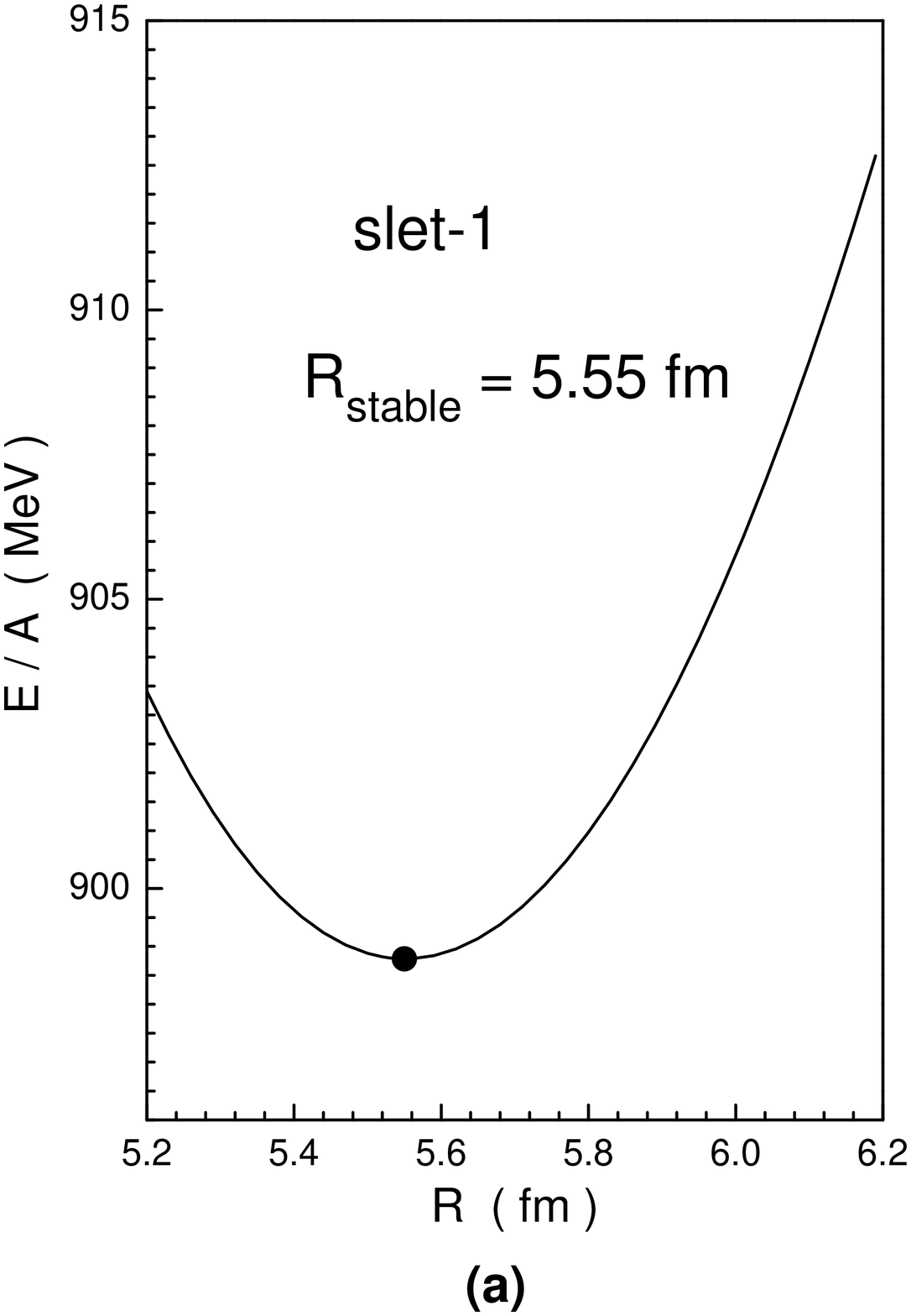,width=4.1cm}
\epsfig{file=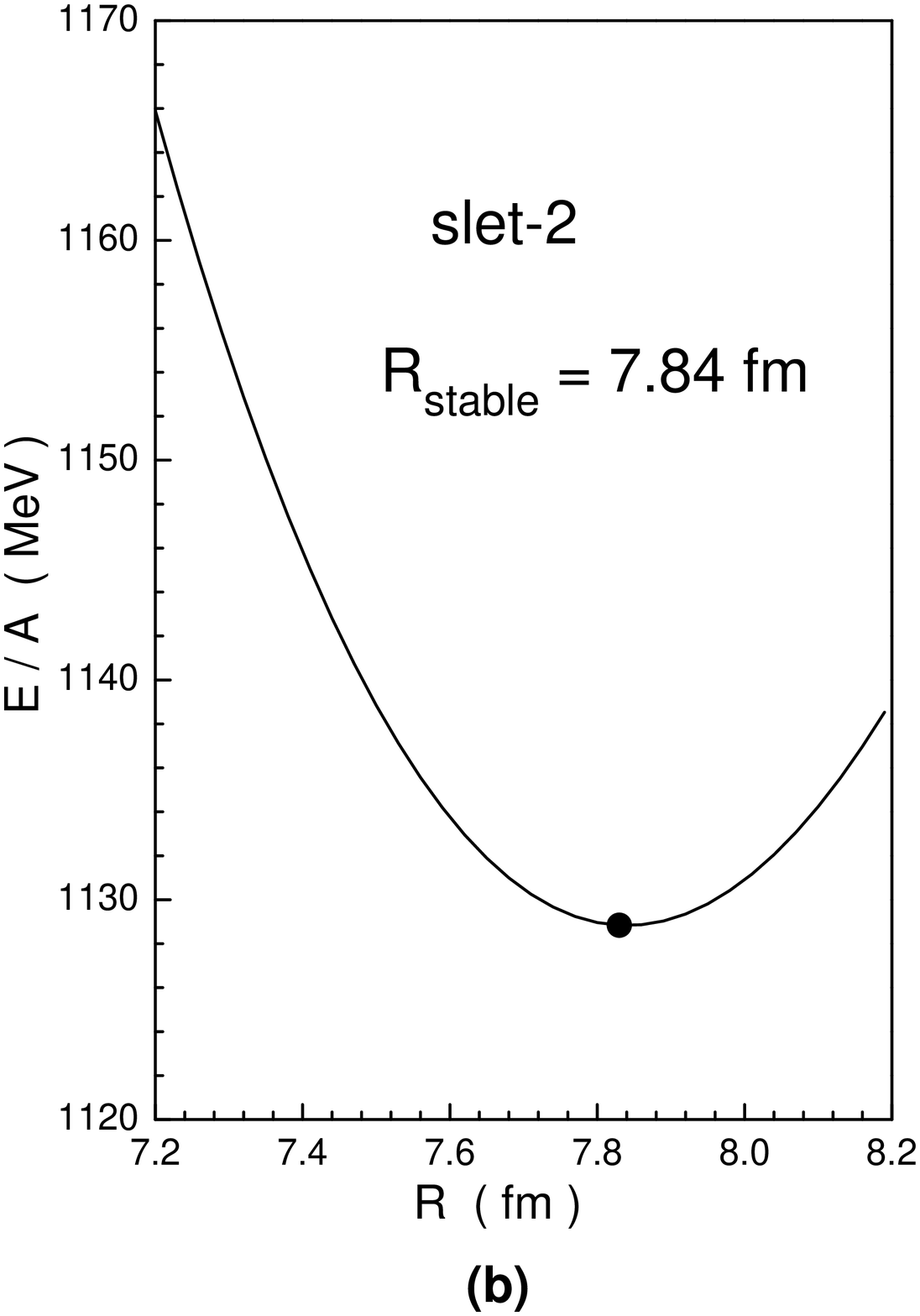,width=4.1cm}
\epsfig{file=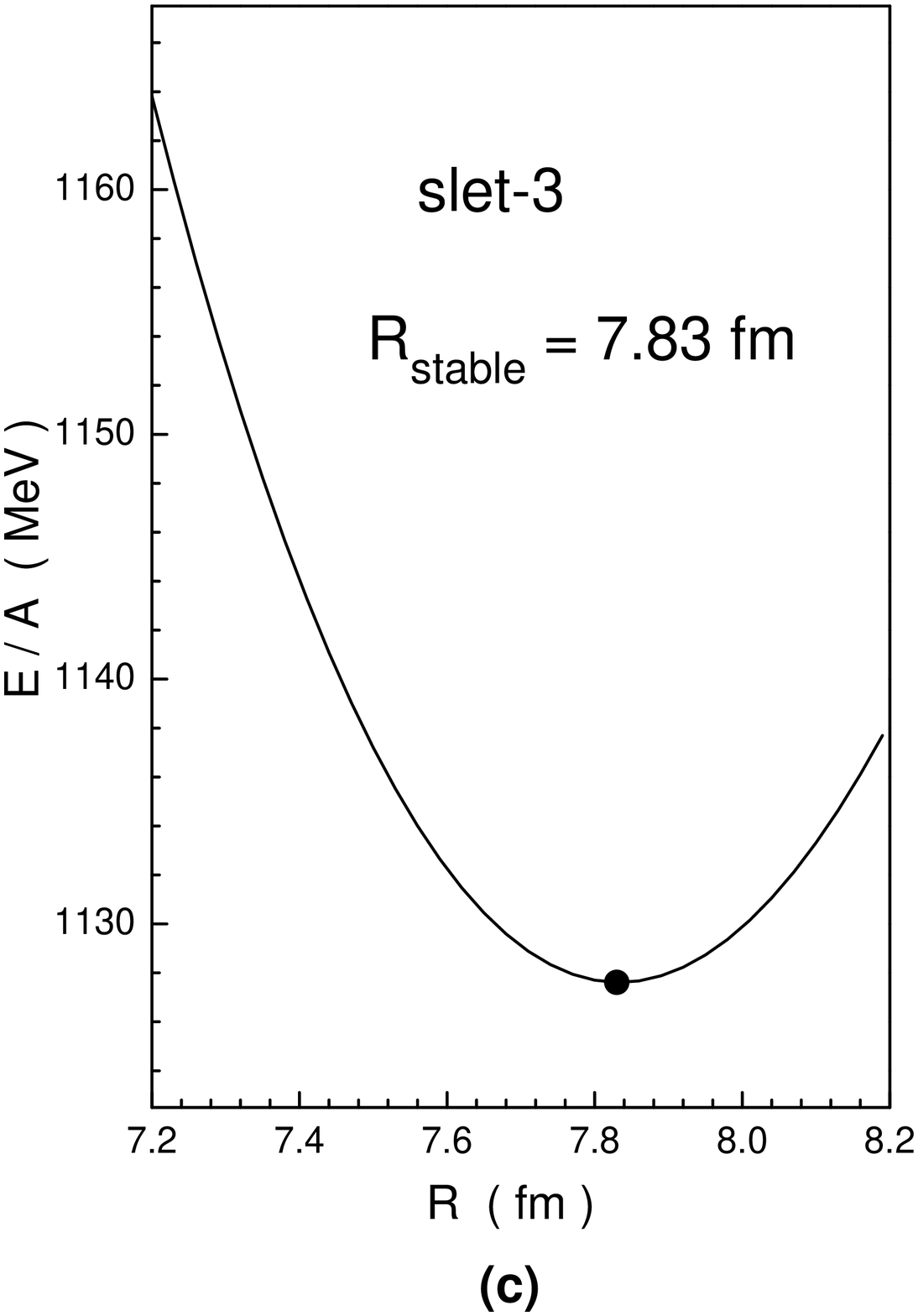,width=4.1cm} \caption{The energy per
baryon of CFL strangelets as functions of the radius. The
mechanically stable radius, marked by full dots and given
numerically by $R_\mathrm{stable}$, is located at the minimum, where
the pressure is exactly zero.  For a fixed baryon number ($A=200$
for the figure), there are three solutions: Figure (a), (b), (c)
respectively for slet-1, slet-2, and slet-3.  \label{fig:EnR}}
\end{figure}

We choose the parameters $B^{1/4}=144$ MeV, $D^{1/2}=120$ MeV in
calculating the data for Figs.~\ref{fig:033eaza},
\ref{fig:slet2slet3} and \ref{fig:EnR}. The energy per baryon and
electric charge per $A^{2/3}$ as a function of baryon number $A$ for
CFL strangelets is given in Fig.~\ref{fig:033eaza}. The strange
current mass $m_{s0}$ is taken to be 80, 120 MeV respectively. We
find the net electric charge $Z\approx0.15A^{2/3}$, which is half of
the result in the pure bag model.\cite{MadsenPRL2001} In the same
figure, the energy of the neutral ordinary strangelets is also
plotted by a dash-dotted line.
When we solve Eq.~(\ref{Eq:chemical}) we find that there are three
different solutions for the common Fermi momentum. In
Fig.~\ref{fig:slet2slet3} the charge of the other two solutions are
given. The charge of CFL slet-2 (dashed lines) is proportional to
cubic-root of baryon number, while CFL slet-3 (dotted lines) is
nearly neutral. The corresponding Fermi momenta are also indicated
on the right axis. Specially, given $A=200$ we can get the smaller
radius $R=5.55$ with $p_\mathrm{F}\approx267$ MeV, which is marked
by slet-1 in Fig.~\ref{fig:EnR}(a). The net positive charge $Z$
satisfies $Z/A^{2/3}\approx0.15$ as mentioned above. The
corresponding high $p_\mathrm{F}$ is the usual case previously
obtained by other authors. With the Fermi momentum decreasing we can
find the other negatively charged solutions marked by slet-2 and
almost neutralized solution marked by slet-3, respectively, in
Fig.~\ref{fig:EnR}(b) and (c). They have larger radius $7.8$ fm. The
corresponding Fermi momenta (42 MeV and 12 MeV or so) are smaller
than that for slet-1 which will be interpreted in the next
paragraph. It can been claimed that slet-1 is more stable at the
same parameters. In Fig.~\ref{fig:EnR}(c) for slet-3, the charge to
baryon number ration is $Z/A\approx-0.0006$, close to neutral state
because $n_u\approx n_d\approx n_s$ is nearly satisfied.

There exist in literature two treatments concerning the deconfined
quark phase. One is the bulk quark matter which is infinite in
volume and without regard to boundary of the sphere. The other is
finite-size lump quark matter, i.e. the so-called strangelets. The
ordinary quark matter can support electrons within it and is charge
neutral in $\beta$ equilibrium. CFL quark matter is automatically
electrically neutral without participation of electrons because
BCS-pairing requires the Fermi momenta of different flavor quarks to
be equal. For ordinary strangelets or CFL strangelets, we should
fall back on how the density of states depend on the quark numbers.
The surface term has a negative contribution to the total number of
massive quarks no matter how large the paring Fermi momentum is. The
curvature term, however, has a positive or negative contribution,
depending on the values of quark mass and Fermi momentum. So
ordinary strangelets is positively charged because the number of
heavier, negatively charged strange quarks is suppressed compared to
the number of lighter up and down quarks. Similarly, the CFL
strangelets (slet-1) is positively charged as in Madsen's work.
\cite{MadsenPRL2001} The Fermi momenta of three cases satisfy
(slet-3)$<$(slet-2)$<$(slet-1). In the case of smaller
$p_\mathrm{F}$, the surface and curvature terms make the number of
strange quarks greater than, or nearly equal to, that of u/d quarks.
As a consequence, the slet-2 is negatively charged. For slet-3, the
very small Fermi momentum make the surface and curvature effect so
weak that slet-3 is nearly neutral. Although there are multiple
solutions for a fixed baryon number, it is necessary to declare that
the positively charged slet-1 is more stable than and the other two
if the density dependence of the pairing parameters is not
considered.

\begin{figure}[pb]
\begin{minipage}[t]{0.48\linewidth}
\includegraphics[width=6.cm,height=6cm]{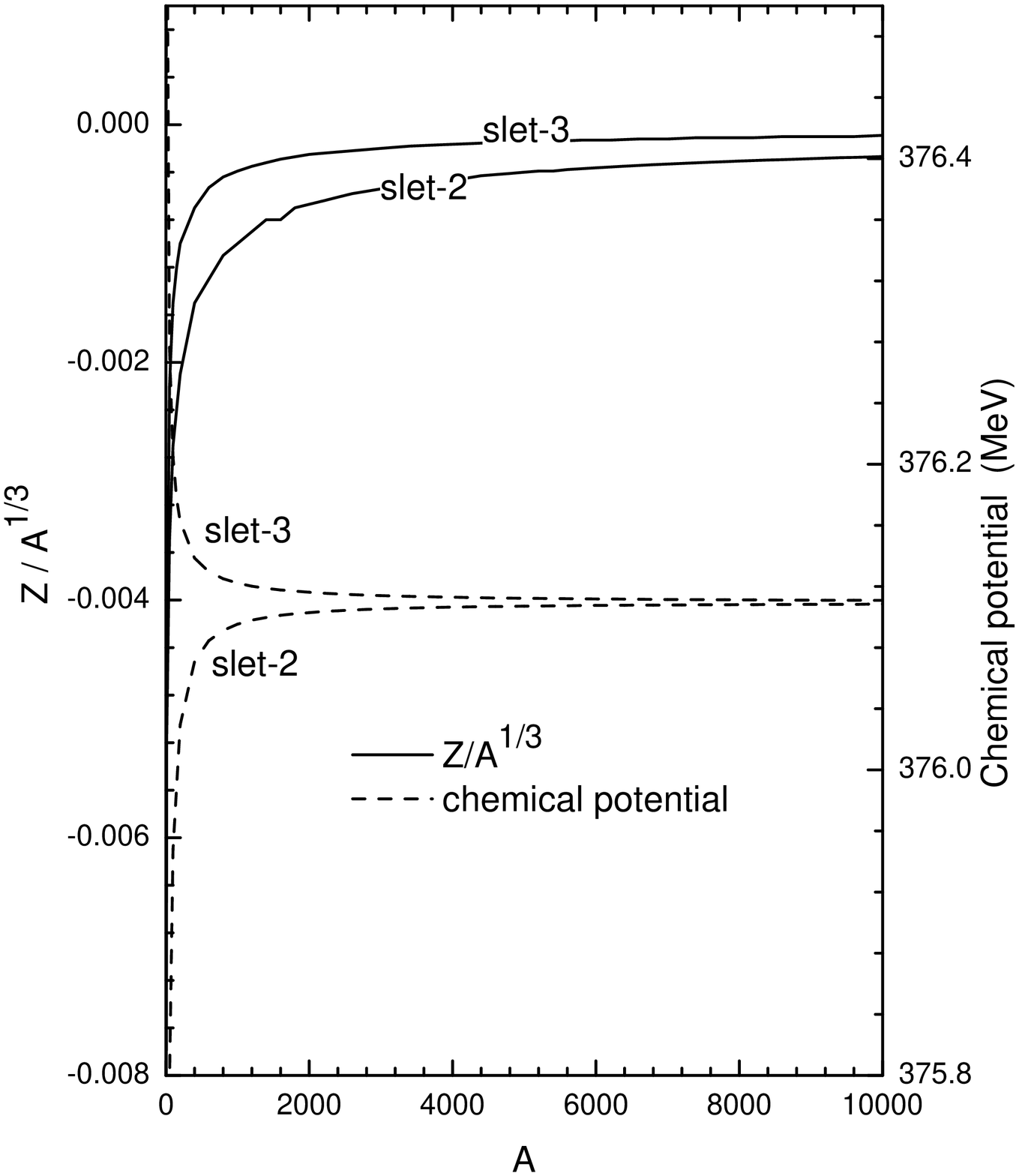}
\caption{The charge to cubic baryon number (solid lines) and
chemical potentials (dashed lines) vs. the baryon number for slet-2
and slet-3.}\label{fig:55}
\end{minipage}
\hfill
\begin{minipage}[t]{0.48\linewidth}
\centering
\includegraphics[width=5.5cm,height=6cm]{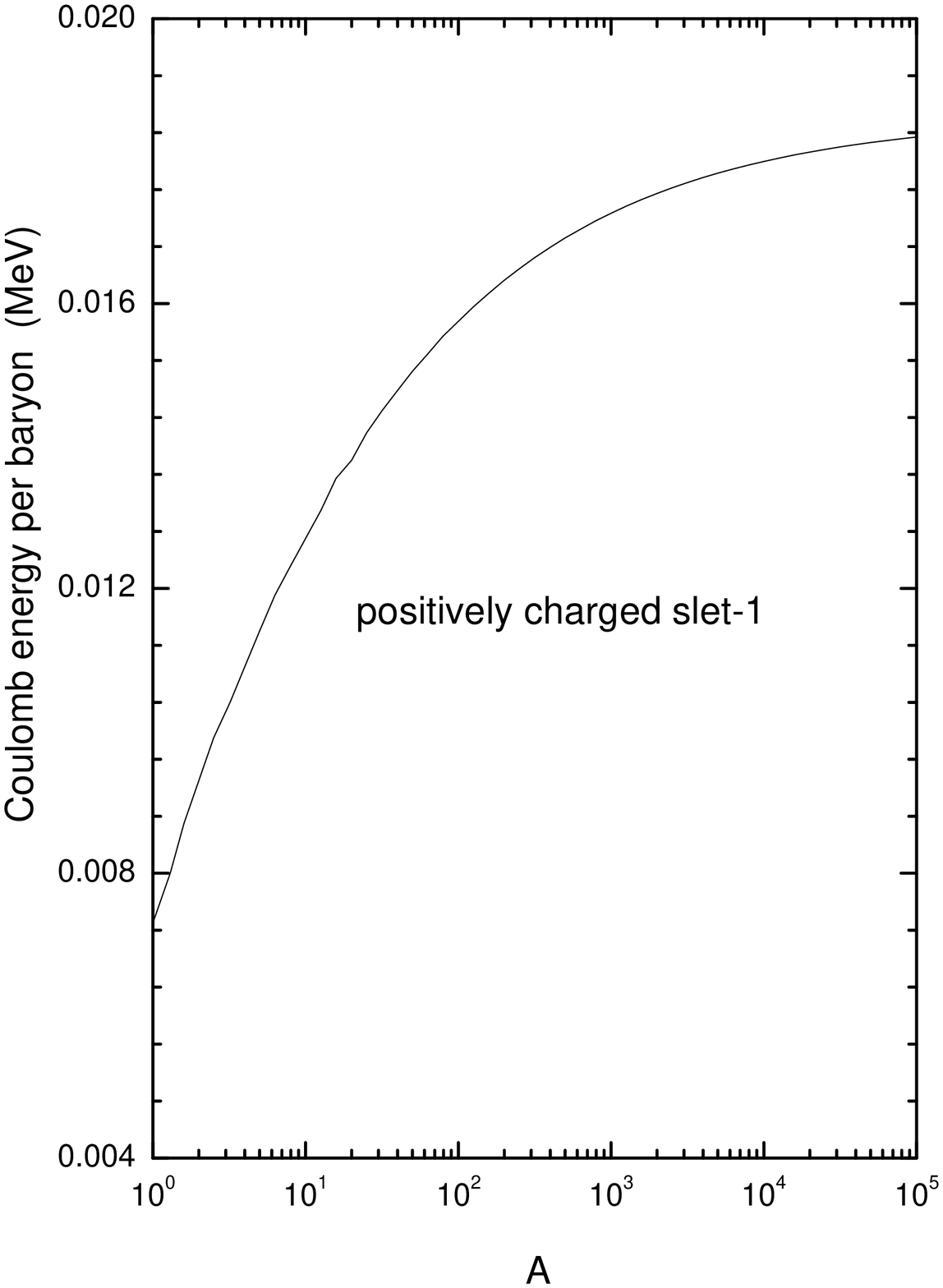}
\caption{The Coulcomb energy per baryon vs. baryon number. Its
effect on the free energy per baryon even will not exceed 0.02 MeV
in our baryon number range. }\label{fig:coul}
\end{minipage}
\end{figure}

It should be emphasized that the common Fermi momentum
$``p_\mathrm{F}"$ is only a fictional intermediate parameter in CFL
matter.  Different from the usual phase, the density of CFL
strangelets depends not only indirectly on the chemical potential
through $p_\mathrm{F}$, but also directly on the chemical potential.
As a consequence the density can be large even if $p_\mathrm{F}$ is
small as long as the chemical potential is big.
Therefore, the density is still higher than, or at most, near the
normal nuclear saturation density, because the chemical potential is
big, though $p_\mathrm{F}$ is small. In fact, the corresponding
chemical potentials $\bar{\mu}$ of slet-2 and slet-3 are larger than
that of slet-1. The results satisfy the validity condition that the
ratio $\Delta/\mu$ is small. So the densities of new CFL strangelets
``slet-2" and ``slet-3" are still in reasonable range. We can still
distinguish them through the difference of the electric charge $Z$
per $A^{1/3}$ and chemical potentials $\bar{\mu}$ in
Fig.~\ref{fig:55}. The smaller the baryon number, the more visible
the difference is. The Coulomb energy of the positively charged
slet-1 has been shown in Fig.~\ref{fig:coul}. With increasing baryon
number $A$, the trend becomes placid. The effect on the free energy
per baryon will not exceed 0.02 MeV. For slet-2 and slet-3 it will
be much smaller than 0.001 MeV.

It should be noted that the new solutions, slet-2 and 3, have an
unusual common momentum. When two paired quarks have a very small
momentum, their global behavior may looks like a boson. Therefore,
the new solution may indicate that boson condensation or diquark
condensate appears to some extent, and the formation mechanism needs
further investigations in the future.

\section{Summary}\label{Sec:conls}

We have studied CFL strangelets within the framework of the quark
mass density-dependent model. We have add the Coulomb interaction to
the charged strangelets. It is found that the positive net charge is
$Z/A^{2/3}\approx 0.15$, nearly half of the previous result in the
pure bag model. Importantly, with decreasing Fermi momentum, we can
find the other two solutions which have larger radius than the
ordinary solution of CFL phase. The new solutions are slightly
negatively charged or nearly neutral. Although with small Fermi
momentum, the strangelets have large chemical potential. Due to the
pairing effect, they have a comparable density with dense matter.
The charge to baryon number of the positively charged strangelets is
smaller than previously found, while that of the negatively charged
strangelets is nearly proportional in magnitude to the cubic-root of
the baryon number. However, present results depend on the parameter
choice, and so, further studies are needed.

\section*{Acknowledgments}

The author would like to thank support from the NSFC (10675137,
10375074, and 10475089).



\begin{thebibliography}{0}    


\bibitem{Rajagopal2000}
K. Rajagopal, F. Wilczek,   At the Frontiers of Particle
Physics/Handbook of QCD, Vol. 3 (Singapore, World Scientific, 2001),
p. 2061-2151.

\bibitem{Alford2001}
M. G. Alford, {\it Annu. Rev. Nucl. Part. Sci.} {\bf 51}, 131
(2001).

\bibitem{JaffePRD1984}
E. Farhi and R. L. Jaffe, {\it Phys. Rev. D} {\bf 30}, 2379 (1984).

\bibitem{BanksPRD2005}
T. Banks, J. D. Mason, D. O'Neil, {\it Phys. Rev. D}  {\bf 72},
043530 (2005).

\bibitem{Khlopov}
D. Fargion, M. Y. Khlopov and C. A. Stephan,
[astro-ph/0511789].
\bibitem{khlopov1}
 M. Y. Khlopov,
[astro-ph/0511796].

\bibitem{Alford1991}
D. Kastor and J. Traschen,
{\it Phys. Rev. D}  {\bf 44}, 3791 (1991).

\bibitem{Alfordjhep}
M. Alford and K. Rajagopal, {\it JHEP} {\bf 0206}, 031 (2002).

\bibitem{Alford2004}
M. Alford, {\it Prog. Theor. Phys. Suppl.} {\bf 153} 1 (2004).

\bibitem{LugonesAA2003}
G. Lugones, J. E. Horvath,
{\it Astron. Astrophys.} {\bf 403}, 173 (2003).

\bibitem{Lugones2004}
G. Lugones, J. E. Horvath, {\it Int. J. Mod. Phys.} {\bf
13}, 1287 (2004).

\bibitem{Lugones2005}
 G. Lugones, I. Bombaci,   {\it Phys. Rev. D}  {\bf 72}, 065021 (2005).

\bibitem{GyulassyNPA2005}
M. Gyulassy, L. McLerran, {\it Nucl. Phys. A} {\bf 750}, 30 (2005).

\bibitem{MAlfordNPB99}
M. G. Alford, K. Rajagopal and F. Wilczek, {\it Nucl. Phys.} {\bf B
537}, 433 (1999).

\bibitem{MadsenPRL2001}
J. Madsen, {\it Phys. Rev. Lett.} {\bf 87}, 172003 (2001).

\bibitem{Huangm03PRD67}
M. Huang, P. Zhuang, and W. Chao,
  {\it Phys. Rev. D}  {\bf 67}, 065015 (2003).

\bibitem{Shovgovy2003}
I. Shovkovy and M. Huang, {\it Phys. Lett. B} {\bf 564}, 205 (2003).

\bibitem{Mishra2005}
A. Mishra and H. Mishra,
  {\it Phys. Rev. D}  {\bf 71}, 074023 (2005).

\bibitem{Kiriyama2005}
O. Kiriyama,
  {\it Phys. Rev. D}  {\bf 72}, 054009 (2005).

\bibitem{Hou2004}
I. Giannakis, D. F Hou, H. C. Ren, and D. H. Rischke, {\it Phys.
Rev. Lett} {\bf 93}, 232301 (2004).

\bibitem{XBZhangPRD2004}
X. B. Zhang and X. Q. Li,  {\it Phys. Rev. D}  {\bf 70}, 014015
(2004).

\bibitem{huazhong}
Y. W. Yu and X. P. Zheng, {\it Astron. Astrophys.} {\bf450}, 1071
(2006).

\bibitem{KRajagopalPRL2001}
K. Rajagopal and F. Wilczek, {\it Phys. Rev. Lett} {\bf 86}, 3492
(2001).

\bibitem{MadsenJPG2002}
J. Madsen, {\it J. Phys. G\/} {\bf 28}, 1737 (2002).


\bibitem{MAlfordPRL04}
M. Alford, C. Kouvaris and K. Rajagopal, {\it Phys. Rev. Lett.} {\bf
92}, 222001 (2004).

\bibitem{MAlfordPRD2005}
M. Alford, C. Kouvaris and K. Rajagopal, {\it Phys. Rev. D} {\bf
71}, 054009 (2005). 

\bibitem{LugonesPRD2002}
G. Lugones and J. E. Horvath,
{\it Phys. Rev. D}  {\bf 66}, 074017 (2002).

\bibitem{henleyNPA1990}
E. M. Henley and H. M¡§uther, {\it Nucl. Phys. A} {\bf 513}, 667
(1990).

\bibitem{BrownPRL1991}
G. E. Brown and M. Rho, {\it Phys. Rev. Lett} {\bf 66}, 2720 (1991).

\bibitem{CohenPRL1991}
T. D. Cohen, R. J. Furnstahl, and D. K. Griegel, {\it Phys. Rev.
Lett} {\bf 67}, 961 (1991).

\bibitem{Cohen1992}
T. D. Cohen, R. J. Furnstahl, and D. K. Griegel, {\it Phys. Rev. C}
{\bf 45}, 1881 (1992).

\bibitem{waleckaNP1995}
J. D. Walecka, {\it Oxford Stud. Nucl. Phys.} {\bf 16}, 1 (1995).

\bibitem{BuballaPLB1999}
M. Buballa and M. Oertel, {\it Phys. Lett. B} {\bf 457}, 261 (1999).


\bibitem{Fowler1981}
G. N. Fowler, S. Raha, R. M. Weiner, {\it Z. Phys. C} {\bf 9}, 271
(1981).

\bibitem{Benrenuto1995}
O. G. Benrenuto and G. Lugones,  {\it Phys. Rev. D}  {\bf 51}, 1989
(1995).

\bibitem{Lugones1995}
G. Lugones and O. G. Benrenuto, {\it Phys. Rev. D} {\bf 52}, 1276
(1995).

\bibitem{PengPRC59}
G.~X.~Peng, H.~C.~Chiang, P. Z. Ning, and B. S. Zou, {\it Phys. Rev.
C} {\bf 59}, 3452 (1999).
\bibitem{Peng00PRC61}
G.~X.~Peng, H.~C.~Chiang, J.~J.~Yang, L.~Li, and B.~Liu, {\it Phys.
Rev. C} {\bf 61}, 015201 (2000).

\bibitem{Peng00PRC62}
G.~X.~Peng, H.~C.~Chiang, B. S. Zou, P. Z. Ning, and S. J. Luo, {\it
Phys. Rev. C} {\bf 62}, 025801 (2000).

\bibitem{Lugones2003qmdd}
G. Lugones and J. E. Horvath,
{\it Int. J. Mod. Phys. D} {\bf 12}, 495 (2003).

\bibitem{Zhang2001}
Y. Zhang, R. K. Su, S. Q. Ying, and P. Wang,
{\it Europhys. Lett.} {\bf 53}, 361 (2001).

\bibitem{Zhang2002}
Y. Zhang, R. K. Su, {\it Phys. Rev. C} {\bf 65}, 035202 (2002);

\bibitem{Zhangprc2003}
Y. Zhang, R. K. Su, 
{\it Phys. Rev. C}  {\bf 67}, 015202 (2003).

\bibitem{Zhang2003}
Y. Zhang, R. K. Su, {\it Mod. Phys. Lett. A} {\bf 18}, 143 (2003).


\bibitem{PRC72(2005)015204}
X. J. Wen, X. H. Zhong, G. X. Peng, P. N. Shen, and P. Z. Ning, {\it
Phys. Rev. C} {\bf 72}, 015204 (2005).

\bibitem{PengPLB2006}
G. X. Peng, X. J. Wen, Y. D. Chen, {\it Phys. Lett. B} {\bf 633},
314 (2006).

\bibitem{MAlfordPRD2001}
M. Alford, K.Rajagopal, S. Reddy, and F. Wilczek,  {\it Phys. Rev.
D}
{\bf 64}, 074017  (2001). 

\bibitem{R.Balian}
R. Balian and C. Bloch, {\it Ann. Phys. (N. Y.)} {\bf 60}, 401
(1970).

\bibitem{Berger1987}
M. S. Berger and R. L. Jaffe, {\it Phys. Rev. C} {\bf 35}, 213
(1987).


\bibitem{Madsen1993}
J. Madsen, {\it Phys. Rev. D} {\bf 70}, 391 (1993).

\bibitem{Madsen1994}
J. Madsen, {\it Phys. Rev. D} {\bf 50}, 3328 (1994).

\bibitem{pertQCD}
E. S. Fraga and P. Romatschke, {\it Phys. Rev. D} {\bf 71}, 105014
(2005).

\bibitem{ScherNPA1997}
K. Schertler, C. Greiner, and M. H. Thoma, {\it Nucl. Phys. A} {\bf
616}, 659 (1997).

\bibitem{lecture2002}
G. Giacomelli, and M. Sioli,  Lectures at the 6th Constantine School
on ``Weak and Strong Interactions Phenomenology¡±, [hep-ex/0211035].

\bibitem{WeberPPNP2005}
F. Weber, {\it Progress in Particle and Nuclear Physics} {\bf 54},
193 (2005).

\bibitem{Rapp1998PRL81}
 R. Rapp, T. Sch\"{a}fer, E. Shuryak, and M. Velkovsky,
{\it Phys. Rev. Lett.} {\bf 81},  53 (1998).

\bibitem{Alford1998}
 M. Alford, K. Rajagopal, and F. Wilczek,
{\it Phys. Lett. B} {\bf 422}, 247 (1998).

\bibitem{Berges1999}
 J. Berges and K. Rajagopal,
{\it Nucl. Phys. B} {\bf 538}, 215 (1999).


\end{thebibliography}
\end{document}